\documentclass[prd,nofootinbib,twocolumn,superscriptaddress,amsmath,amssymb]{revtex4}
\usepackage{graphics}
\usepackage{color}
\usepackage{bm}

\begin{document}

\title{Edge magnetoplasmons in a partially screened two-dimensional electron gas on a helium surface}

\author{M. I. Goksu}
\address{Department of Physics, Case Western Reserve University, Cleveland, Ohio 44106-7079, USA}
\address{Truman State University Science Division, 100 East Normal Street, Kirksville,
 Missouri 63501 USA}
\author{Mokyang R. Kim}
\altaffiliation[Present address:]{ Department of Physics, California State
University-Dominguez Hills, Carson, CA 90747, USA}
\author{K. A. Mantey}
\altaffiliation[Present address:]{ Department of Physics, University of Illinois,
Urbana, IL 61801-3080, USA}
\author{A. J. Dahm}
\address{Department of Physics, Case Western Reserve University, Cleveland, Ohio 44106-7079, USA}
\date{October 23, 2006}

\begin{abstract}
\noindent We report a study of edge magnetoplasmons in a partially-screened system of
electrons on a helium surface.  We compare experimental results with theories of the
frequency, damping, and penetration-depth dependence on magnetic field,
temperature-dependent damping, and the dependence of the frequency on screening.  We
show explicitly the dependence of frequency on the edge density profile.  The frequency
and screening are in qualitative agreement with the theory of Fetter at small fields,
and the frequencies agree with theory in the limit of zero magnetic field. The frequency
and linewidths in intermediate and large fields exhibit the features of the qualitative
predictions of Volkov and Mikhailov, but differ numerically. Deviations from theory for
a finite sample occur at smaller fields.  The dependence of frequency on the density
profile is stronger than predicted by these authors, and the penetration-depth variation
with field confirms their prediction for small fields.


\end{abstract}

\maketitle

\section{Introduction}
Edge magnetoplasmons (EMPs) are plasma modes that propagate around the perimeter of a
two-dimensional ($2D$) array of charges.  A degenerate mode at zero magnetic field
splits into two modes at finite B fields.  The frequency of one mode increases with
field, while the frequency of the EMP mode decreases.  The penetration depth of the
upper mode increases with field and becomes infinite at a finite value of B, thereby
becoming a bulk $2D$ mode.  The penetration depth of the lower mode decreases with
increasing B.

Edge magnetoplasmons were first detected in a partially-screened array of electrons on a
helium surface by Mast et al. \cite{Mast,Mast2} and by Glattli et al.\cite{Glattli}
Screening is provided by metallic plates located above and below the electron layer and
is described by a screening length, which depends on the separation of the sample from
these plates.  Both groups developed theories, with which they compared their data.
Earlier modes observed in a quantum dot\cite{Allen} have since been interpreted as EMP
modes. Edge modes have been investigated extensively in 2D semiconductor samples.

Glattli et al.\cite{Glattli2} extended their studies to low and intermediate magnetic
fields. Peters et al.\cite{Peters} and Monarkha et al.\cite{Ito} measured the damping of
these modes at high magnetic fields. Studies of EMPs at a density discontinuity have
been studied by Sommerfeld et al.\cite{Sommerfeld1} Experimental studies of related
acoustic edge modes with $n$ radial nodes ($n>1$) have been made on
electron\cite{Kirichek1,Sommerfeld2,Kirichek2} and ionic\cite{Elliot} arrays at the
surface of liquid helium. Edge modes in a sample of electrons on a helium surface have
been treated theoretically by a number of
authors.\cite{Fetter1,Fetter2,Fetter,Volkov,Mikhailov,VM,Nazin1,Nazin2} Here we present
a study of EMP modes in a partially screened system in a circular sample for small and
intermediate fields and compare them with theories by Fetter\cite{Fetter} and Volkov and
Mikhailov.\cite{Volkov,Mikhailov} Preliminary work has been presented
elsewhere.\cite{Goksu1,Goksu2}

\section{Theory}
Fetter\cite{Fetter} solved for the edgemagnetoplasmon modes of a
2D electron array in a circular sample of radius $R$ located
symmetrically between two metallic plates each located a distance
$h$ from the electron layer.  A magnetic field is applied along the
$z$ axis normal to the layer.  A step density profile at the sample
perimeter was assumed. He solved a hydrodynamic model of a
compressible charged fluid placed in a uniform neutralizing
background.  The continuity and Euler equations that describe the
dynamics are
\begin{equation}
\partial n_\ell/\partial t + n\nabla\cdot\overrightarrow{v} = 0,
\end{equation}
\begin{equation}
\partial \overrightarrow{v}/\partial t +c^2n^{-1}\nabla n_\ell -(e/m)\nabla\Phi
-\omega_c\overrightarrow{z}\times\overrightarrow{v} = 0,
\end{equation}
where $n_\ell$ is the local electron density, $n$ is the average
electron density, $\overrightarrow{v}$ is the local velocity of the
electron fluid, $c$ is the effective wave speed that allows for
dispersion in the propagating wave, $\Phi$ is the electrostatic
potential at the plane of the charges $(z = 0)$, $\omega_c$ is the
cyclotron frequency, and $\nabla$ is the two-dimensional gradient
operator. Poisson's equation relates the potential and the charge
density;
\begin{equation}
\nabla^2\Phi = n_\ell\delta(z)/\epsilon_0.
\end{equation}

A solution of the form $e^{-(\overrightarrow{q}\cdot\overrightarrow{r}-\omega t)}$,
where $\overrightarrow{q}$ lies in the $xy$ plane, is substituted into these equations
and boundary conditions are applied.  With a Hankel transform of the potential and
neglect of the sound velocity for long wavelengths, Fetter arrives at the following
equations to be solved for the resonance frequencies.
\begin{equation}
\Sigma_{j=0}^\infty[K_{ij}-(\frac{\omega}{\Omega_0}-\frac{\omega_c}{\Omega_0})^2\gamma_{ij}-(\frac{\omega^2}{\Omega_0^2}-\frac{\omega_c^2}{\Omega_0^2})g_{ij}]c_j=0.
\end{equation}
Here,
\begin{equation}
\Omega_0^2 = \frac{ne^2\tanh(h/R)}{mR\epsilon_0(\epsilon +1)},
\end{equation}
where $\epsilon$ is the dielectric constant of liquid helium.  The
Kernel matrix $K_{ij}$ is defined as
\begin{equation}
K_{ij} = \coth(\frac{h}{R})\int \frac{dp}{p^2}
\tanh(\frac{ph}{R})J_{L+2i+1}(p)J_{L+2j+1}(p),
\end{equation}
where $J(p)$ is a Bessel function, and $L=|k|$ is an integer; $k$ is the azimuthal mode
number. The matrix $\gamma_{ij}$ has only the single nonzero element
\begin{equation}
\gamma_{00} = [8L(L+1)^2]^{-1}.
\end{equation}
The matrix element $g_{ij}$ is symmetric and tridiagonal with nonzero elements,
\begin{equation}
g_{ii} = [4(L+2i)(L+2i+1)(L+2i+2)]^{-1},
\end{equation}
\begin{equation}
g_{i,i+1} = [8(L+2i+1)(L+2i+2)(L+2i+3)]^{-1}.
\end{equation}

Volkov and Mikhailov\cite{Volkov,Mikhailov} used the Weiner-Hopf method to solve Eqs.
(1)-(3). These authors defined the penetration depth $\lambda$ of the EMP mode into the
sample as the largest of $h$, the width of the density profile $b$, and a magnetic
length $\ell$. For small $qh$, $\omega^2 \ll \omega_c^2$, and $\omega\tau \gg 1$, $\ell$
is given by
\begin{equation}
\ell=\frac{ne^2}{m\epsilon_0(\epsilon+1)\omega_c^2}.
\end{equation}
Here q is the azimuthal wave vector, and $\tau$ is the scattering time for B = 0.
Since the lengths $h$ and $b$ are nearly equal for our sample, we take for our
analysis
\begin{equation}
\lambda=(\ell^2 + b^2)^{1/2}.
\end{equation}

For the circular geometry used in this experiment these authors find
\begin{eqnarray}
\omega_L = -\frac{\alpha \sigma_{xy}q}{\pi\epsilon_0(\epsilon+1)} = \frac{\alpha
\textit{neL}}{\pi\epsilon_0(\epsilon+1)\textit{RB}},\\
\nonumber \alpha = \ln\frac{2R}{\lambda}-\Psi(L+\frac{1}{2})+1,
\end{eqnarray}
where $\sigma_{xy}$ is the Hall conductivity, and $\Psi$ is the digamma function. This
formula was derived for a step density profile and $\ell \ll \pi R$, which for our
system requires $B \gg 10^{-3}$ T. For a finite density-profile width we have
substituted $\lambda$ for $\ell$ in their argument of the logarithm.
\begin{figure}[t]
\begin{center}
\scalebox{0.50}{\includegraphics{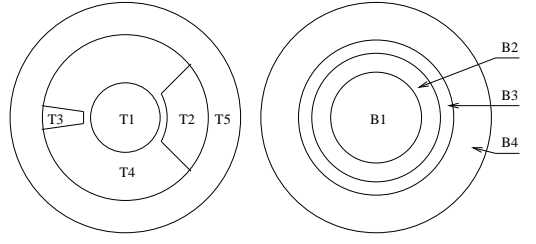}} \caption{Configuration of electrodes
located above and below the electron layer. The electrodes in the top $T$ and bottom $B$
plates are described in the text.}
\end{center}
\end{figure}

These authors also obtain a formula that includes the density profile $\Theta(x/h)$ for
a semi-infinite plane, where $x$ is measured from the edge of the sample. Adapting their
formula to a circular geometry as an approximation by setting $q=L/R$, the frequency is
obtained with a screening length of $h/2$ as
\begin{eqnarray}
\omega_L = -\frac{\beta \sigma_{xy}q}{\pi\epsilon_0(\epsilon+1)} = \frac{\beta
\textit{neL}}{\pi\textit{RB}\epsilon_0(\epsilon+1)},\\
\nonumber \beta = [\ln \frac{2h}{\pi b}+A],\\
\nonumber A = \int d\xi \ln (\frac {1}{\xi}) \frac{\partial\Theta(\xi)}{\partial \xi}.
\end{eqnarray}
This equation was derived under the conditions $h/R\ll 1$ and $b\ll h$.

Shikin's and Nazin \cite{Nazin2} included the density profile by solving Eqs. (1)-(3)
numerically for the case of specific density profiles in a circular geometry.

Volkov and Mikhailov\cite{Mikhailov} derived the linewidth of the modes
for a semi-infinite plane and $\omega\tau >> 1$.  We adapt their formula by
substituting $q = L/R$;
\begin{equation}
\Delta\omega\approx\frac{1}{\omega_{L}\tau} \frac{neL}{\epsilon_0(\epsilon+1)RB}.
\end{equation}

\section{Experimental Setup}
Our sample cell consists of a plane-parallel circular capacitor with a plate separation
of $2$ mm.  Electrons are deposited on a helium surface located midway between the
plates. The capacitor plates are made of copper-clad epoxy board, and electrodes are
etched on the inner side of each plate.  Voltages are applied on these electrodes to
excite EMP modes and to fix the radius and density of the electron pool.

The electrode configuration is shown in Fig. 1. The bottom plate is separated into four
concentric electrodes of outer radii $5.8$, $8.2$, $10$, and $13$ mm.  The top plate
includes three electrodes on which a $rf$ voltage is applied to drive resonant modes.
The inner circular electrode $T_1$ is used for radial modes, and two electrodes $T_2$
and $T_3$ spanning angles of $80^\circ$ and $12^\circ$, respectively, are used to drive
EMP modes. A larger electrode $T_4$ is used in conjunction with electrodes in the lower
plate as a capacitor plate to measure the helium level. A guard voltage $V_g$ is applied
to electrodes, $T_5$ and $B_4$, and to a $2$ mm high circular electrode just beyond
these to confine the sample and shape the electron density profile. For smaller pool
radii the guard voltage is also applied to one or two of the inner circular electrodes
in the bottom plate.

We work at the saturated density $n=V_h/\epsilon\epsilon_0 eh$. The melting temperature
serves as a check on the density and helium level.  We estimate the density values to be
accurate to within $5\%$.
\begin{figure}[t]
\begin{center}
\resizebox{7cm}{!}{\rotatebox{90}{\includegraphics{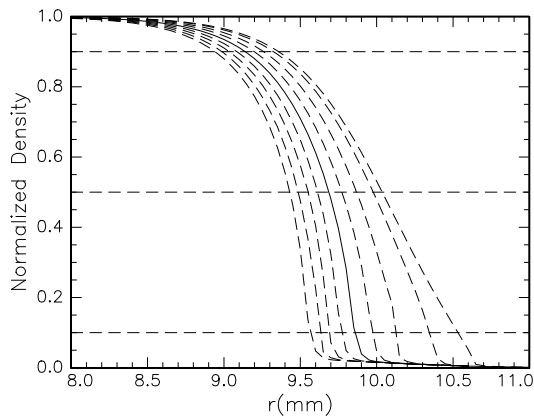}}} \caption{Density
profiles for different ratios of guard to holding voltages.  The ratios of $|V_g|/V_h$
are from left to right $0.8, 0.7, 0.6, 0.5, 0.4, 0.3, 0.2, 0.1, 0.05$}.
\end{center}
\end{figure}

A broadband, homodyne, rf reflection spectrometer is used to detect the power absorbed
as a function of frequency. A $50$ $\Omega$ coax line is terminated in a $50$ $\Omega$
resistor in parallel with the driving electrode.  A line of equivalent electrical length
is added to the reference arm of the spectrometer. Both arms of the spectrometer have
identical components so that there is no relative phase shift as the frequency is swept.
The electron density and the radius of the pool are modulated by an audio-frequency
voltage applied to the guard electrodes for phase-sensitive detection.  The cell is
leveled by maximizing the steepness of the mobility change at the melting transition of
the sample.

The radius and width of the sample profile are determined by a numerical simulation.
Profiles are shown in Fig. 2 for various ratios of $|V_g|/V_h$.  We take the radius of
the sample $R$ to be the value at which the density profile extrapolates to zero. We
define the width of the profile, $b_{s}$, as the separation of the points where the
density is $10\%$ and $90\%$ of the density at the center. The parameters are
$|V_g|/V_h=0.5$ and $b_{s} = 0.7$ mm for data presented in Figs. 3, 4, 8, and 9, and
$|V_g|/V_h=0.8$ and $b_{s} = 0.6$ mm for data presented in Figs. 5, 6, and 10.  Data are
taken between $300$ and $400$ mK unless indicated.  Only the linewidth data are
temperature dependent.
\begin{figure}[t]
\begin{center}
\resizebox{8.3cm}{!}{\rotatebox{90}{\includegraphics{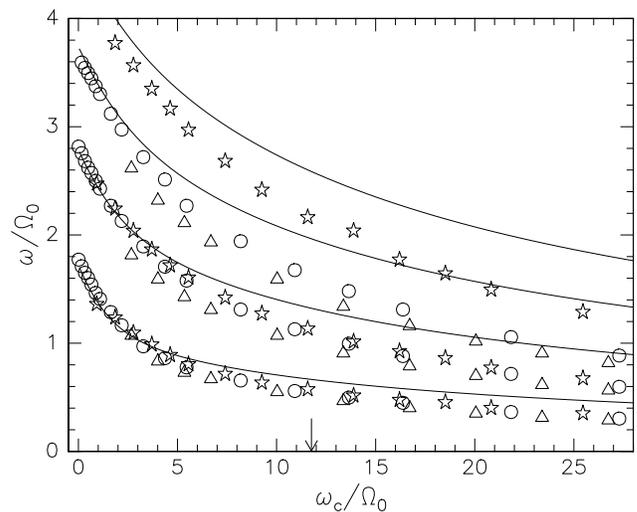}}} \caption{Normalized mode
frequencies versus normalized cyclotron frequency.  $R=9.8$ mm.  Azimuthal modes
$L=1-3$: triangles, $n_{12}=0.7$; circles, $n_{12}=1.05$. $L=1,2$, and $4$: stars,
$n_{12}=1.46$. $\Omega_{0}/2\pi=20.2(n_{12})^{1/2} MHz$.  Frequency and field values
range from $5-92$ MHz and $0-0.022$ T, respectively. Curves represent Fetter's theory
for $L=1-4$.}
\end{center}
\end{figure}

\section{Results}
\subsection{Mode frequencies}

In the following graphs arrows on the abscissa indicate the value of $\omega_c/\Omega_0$
or magnetic field at which $\ell=b_s$.  The frequencies probed range from $2.6$ to $92$
MHz.
\begin{figure}[t]
\begin{center}
\resizebox{8.3cm}{!}{\rotatebox{90}{\includegraphics{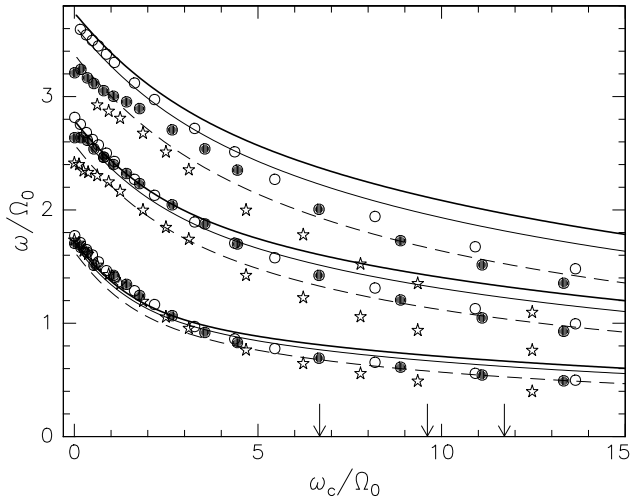}}} \caption{Normalized mode
frequencies versus normalized cyclotron frequency for $n_{12}=1.05$, $L=1-3$, and three
ratios of $h/R$.  Open circles, $h/R=0.10$; closed circles, $h/R=0.13$; stars,
$h/R=0.18$. $\Omega_{0}/2\pi = 203/R(mm) MHz$, with $R$ values of $9.8, 8$, and $5.6$
mm.  The frequency range is $10-74$ MHz.  The maximum field is $0.016$ T.  Curves
represent Fetter's theory: heavy solid, $h/R=0.10$; light solid, $h/R=0.13$; dashed,
$h/R = 0.18$.}
\end{center}
\end{figure}

The data at small magnetic fields are compared with Fetter's theory\cite{Fetter} in
Figs. $3$ and $4$. The kernel $K_{ij}$ given in Eq. (6) depends on the screening ratio
$h/R$ but is independent of density. It follows from Eq. (4) that, within the theory of
Fetter, for a fixed value of $h/R$ a normalized plot of $\omega/\Omega_0$ versus
$\omega_c/\Omega_0$ is universal. In Fig. $3$ normalized data are compared with Fetter's
theory for three densities given in units of $n_{12}=10^{12}$ $m^{-2}$.  Level crossings
with modes of the upper branches make it difficult to determine the frequency of EMP
modes at small fields for larger $L$ values.  The $L=3$ and $4$ modes are weakly coupled
and missing for some densities.  Fetter's theory is plotted for the four lowest EMP
modes using a $25\times 25$ determinant to solve Eq. (4).

The frequency dependence for three ratios of $h/R$ is shown in Fig.
4. The theory of Fetter is shown using a $25\times 25$ determinant
in evaluating Eq. (4) for $h/R=0.10$ and a $15\times15$ determinant
for $h/R=0.13$ and $0.18$.

Data are first taken for $h/R= 0.10$ with the saturated sample above electrodes $B_{1}$,
$B_{2}$, and $B_{3}$.  Then repelling voltages are sequentially applied to electrodes
$B_{2}$ and $B_{3}$ for data sets at $h/R=0.13$ and $0.18$ to insure that the density is
the same for the three sets.

Data for a density of $n_{12} = 0.7$ are compared with the theories of Volkov and
Mikhailov\cite{Mikhailov} for larger magnetic fields in Figs. 5 and 6.  Figure 5 is a
plot of frequency versus magnetic field.  Only data for $L = 1$ and even modes are shown
for clarity.  Likewise, theoretical curves, Eq. (12) with $b = 0.6$ mm and Eq. (13), are
shown only for modes $1, 2, 6$, and $12$. These curves are multiplied by a factor of
$0.4$. Equation (12) with $b = 0$ multiplied by a factor of $0.25$ is shown for $L = 1$
and $12$.

\begin{figure}[t]
\begin{center}
\resizebox{8.3cm}{!}{\rotatebox{90}{\includegraphics{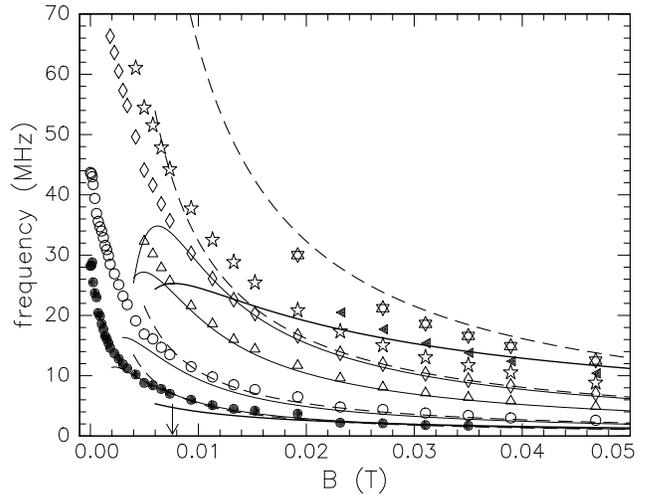}}} \caption{Frequency
versus magnetic field.  $n_{12}=0.7$.  $R=9.6$ mm.  Data are for $L=1$ and even modes
from $L=2$ to $12$.  Curves represent the theories of Volkov and Mikhailov multiplied by
a factor $\kappa$.  For modes $L = 1,2,6$, and $12$ with $\kappa= 0.4$: fine solid, Eq.
(12) with $b=0.6$ mm; dashed curve, Eq. (13).  Bold solid, Eq. (12) with $b=0$,
$\kappa=0.25$ for $L=1$ and $12$.}
\end{center}
\end{figure}

\begin{figure}[b]
\begin{center}
\resizebox{8.3cm}{!}{\rotatebox{90}{\includegraphics{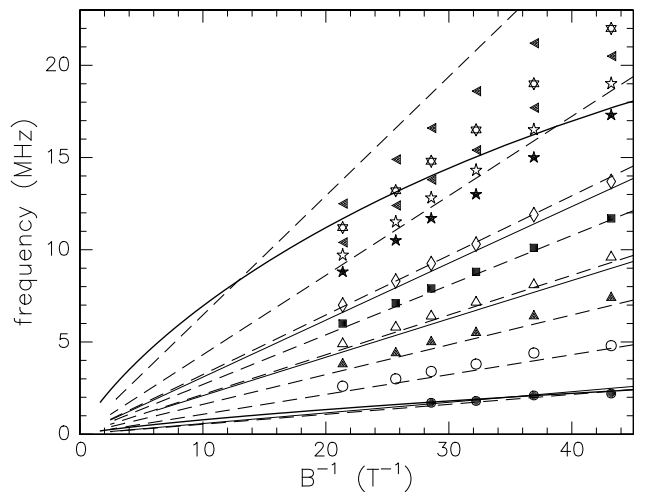}}} \caption{High field
data plotted as frequency versus inverse field. $n_{12}=0.7$. $R=9.6$ mm.  Symbols are
for modes $L=1-6, 8-12$. Theories curves are multiplied by a factor $\kappa$. Light
solid, Eq. (12), $b=0.6$ mm, $L=1,6$, and $12$; dashed, Eq. (13) for $L=1-6, 8$, and
$12$; bold solid, $b = 0$ for $L = 1$ and $12$; The factors $\kappa$ are as labeled in
Fig. 5.}
\end{center}
\end{figure}
In Fig. 6 the high-field data are plotted as a function of inverse field in the range
$\ell\ll b$, where theory predicts $f \propto B^{-1}$.  The lowest $12$ azimuthal modes
are shown with the exception of $L=7$.  This mode couples weakly, since the $rf$ driving
electrode $T_{2}$ spans approximately $1.5$ wavelengths for this mode. Theoretical
curves are shown with the same multiplicative factors used in Fig. 5.

We show the dependence of frequency on the density profile in Fig. 7 for three magnetic
field strengths with the sample confined above electrode $B_{1}$. The mode frequency is
plotted versus $|V_{g}|/V_{h}$.  The calculated profile width $b_{s}$, is given on the
top axis.  This scale is inaccurate below $b_{s} = 0.66$  mm.  The solid curve shows the
variation of $\alpha/R$ from Eq. (12) with $\lambda = b_{s}$.

There was no detectable change in the mode frequencies, less than $1\%$, upon crossing
the melting curve.

\subsection{Linewidths}
The signal profiles were Lorentzian, which shows that the modes are not parametrically
driven. An example of the derivative signal is shown in the inset of Fig. 8.

The linewidths $\Delta\omega/2\pi$ for three azimuthal quantum numbers are plotted as a
function of B in Fig. 8.  The theoretical linewidths have only a weak dependence on L
through the digamma function in the expression for $\omega_{L}$, Eq. (12). The filled
symbols represent Eq. (14) for $L=1$ with experimental values of the angular
frequencies.  The value of $\tau$ is taken from the theory of Vi'lk and
Monarkha.\cite{Vilk} Theoretical values of the linewidths are divided by a factor of $9$
to fit the data. The $Q$'s of the resonances are approximately $\omega/\Delta\omega = 5,
9$ and $17$ for modes $L = 1,2$, and $4$, respectively. There is a small increase of
about $20\%$ with field over the range plotted in Fig. 8.
\begin{figure}[t]
\begin{center}
\resizebox{7cm}{!}{\rotatebox{90}{\includegraphics{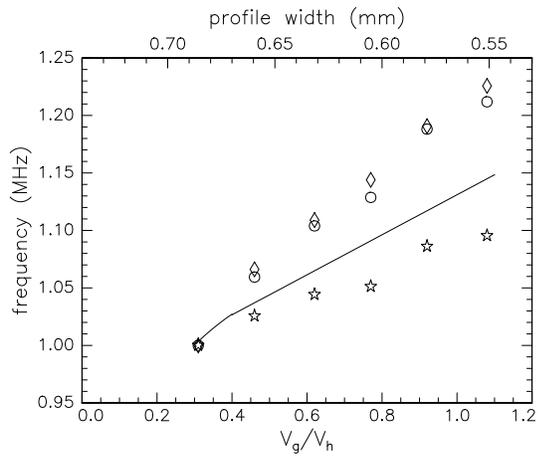}}} \caption{Frequency versus
$|V_{g}|/V_{h}$.  $n_{12}=0.7$. $L=1$. Stars, $B=0$; diamonds, $B=0.0237 T$; circles,
$B=0.0474 T$. Solid curve represents Eq. (12).  Both data and theory are normalized to
the value at $|V_{g}|/V_{h}=0.3$}
\end{center}
\end{figure}

\begin{figure}[b]
\begin{center}
\resizebox{7cm}{!}{\rotatebox{90}{\includegraphics{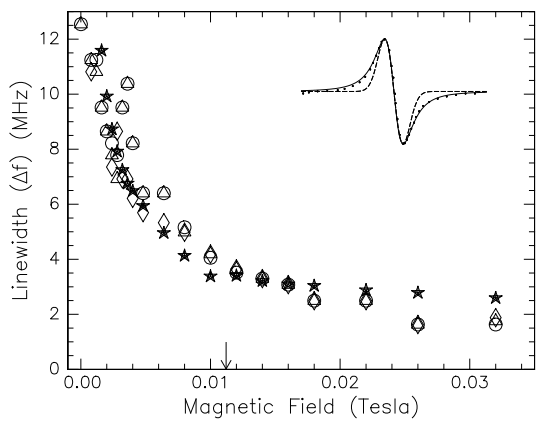}}} \caption{Linewidth
versus magnetic field; $T=400$ mK, $n_{12}=1.75$, $R=8$ mm. Symbols: circles, $L=1$;
triangles, $L=2$; diamonds, $L=4$. Filled symbols represent Eq. (14) for $L=1$ with
experimental values of $\omega_{L}$. Here theoretical values are divided by a factor of
$9$. Inset: Signal profile. The solid trace is a Lorentzian fit; the dashed trace is a
Gaussian fit.}
\end{center}
\end{figure}
Linewidths are shown as a function of temperature in Fig. 9 and compared with theory.
Values are included for the four lowest modes. The curves are given for $L = 1$ by Eq.
(14) divided by a factor of $3$ for $32$ and $160$ G with the experimental value of
$\omega_{L}$ and the theoretical expression\cite{Vilk} for the ripplon-scattering
collision time.

\subsection{Penetration depth}

The variation of the penetration depth of the EMP modes with magnetic field can be
extracted from the data. The area under the Lorentzian power absorption versus frequency
curve is $\pi P_0\Delta\omega$, where $P_0$ is the maximum at the resonant frequency.
The area is proportional to the number $N$ of electrons that participate in the EMP mode
and independent of the collision time. The effective number of electrons is
\begin{equation}
N = 2\pi R\int d\xi \;ne^{(-\xi/\lambda)}\;\; \propto\;\; \lambda.
\end{equation}
Here $\xi$ is the coordinate measured from the edge of the sample
with an assumed step profile, and $\lambda < R$ is assumed. The
magnitude of the extrema in the derivative signal $S$ is proportional to
$P_0/\Delta\omega$. The penetration depth dependence on field is
then
\begin{equation}
\lambda(B) \propto S(\Delta\omega)^2.
\end{equation}

\begin{figure}[t]
\begin{center}
\resizebox{7cm}{!}{\rotatebox{90}{\includegraphics{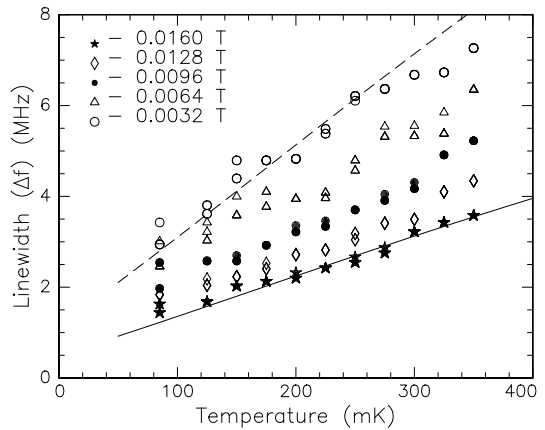}}} \caption{Linewidth
versus temperature for different values of B. $n_{12}=0.88$. $R=9.6$ mm.  Identical
symbols are used for the lowest four modes. Theoretical curves as explained in the text
are dashed line, $32$ G; solid line, $160$ G.}
\end{center}
\end{figure}

This product $S(\Delta\omega)^2$ is independent of temperature for fixed magnetic field
to within the scatter of the data. The product $S(\Delta\omega)^2$ is plotted as a
function of magnetic field in Fig. 10. The curves represents $\lambda =(\ell^2 +
b^2)^{1/2}$; solid - $b = 0.6$ mm, dashed - $b=0$.

\section{Discussion}

Our data are in close agreement with the theory of Fetter at zero field as observed
in Fig. 3. For finite fields the data deviate from theory by an
amount that increases with magnetic field.  The deviation increases
with mode number.  For data taken at $n_{12} = 0.7$, deviation from the two other
normalized experimental curves may be due to a loss of electrons.

Fetter's theory assumes a step density profile at the sample edge. Thus it is applicable
only for $\ell\ll b$, where $\ell\propto (\omega_{c}/\Omega_{0})^{-2}$. The fit of
theory to the data is good for the lowest two modes for $\ell < 10 b$ or
$\omega_{c}/\Omega_{0}<3$ in Fig. 3.  We believe the discrepancy between experiment and
theory is not caused by using too few terms in evaluating Eq. (4). The solutions to Eq.
(4) using $15\times15$ and $25\times 25$ determinants differed by about $1\%$ at
$\omega_{c}/\Omega_{0} = 15$.  Unfortunately, the range in which we can test this theory
is restricted.

The dependence of frequency on the screening parameter $h/R$, shown in Fig. 4, is in
qualitative agreement with theory, and the separation of curves for different values of
$h/R$ is comparable to that predicted by Fetter. Arrows on the curves indicate the
values of $\omega_{c}/\Omega_{0}$ at which $\ell = b_{s}$ with smaller values
corresponding to larger ratios of $h/R$. The reduced slope for larger values of $h/R$
and $L=2$ and $3$ at small fields is not a feature for most of our data.  It may have
resulted from an error in identifying the exact frequency for modes $L$ due to overlap
with the upper-branch modes of $L - 1$.

The theory of Volkov and Mikhailov, Eq. (12), is compared with the field dependency of
our resonant frequencies in the range of applicability, $B\gg 10^{-3}$ T, in Figs. 5 and
6. The theory gives the general qualitative behavior of the data.  However, the digamma
function in the coefficient $\alpha$ overly confines the separation between adjacent
azimuthal modes. It also gives a decrease in frequency. in contradiction with
experiment, for $\ell < b_{s}$, although this is not a condition for the applicability
of this formula.

\begin{figure}[t]
\begin{center}
\resizebox{7cm}{!}{\rotatebox{90}{\includegraphics{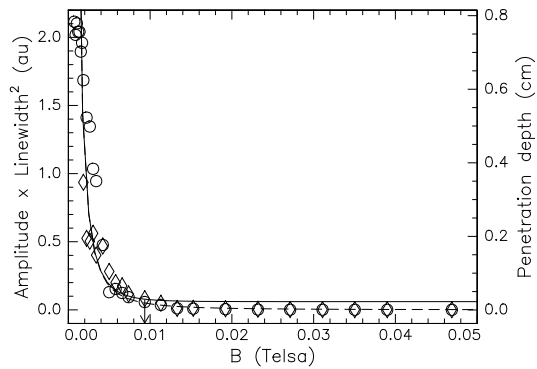}}} \caption{The product
$S(\Delta\omega)^{2}$ in arbitrary units versus magnetic field. $n_{12}=1.05$. $R=9.6$
mm.  Circles, $L=2$; diamonds, $L=5$. The amplitude for each mode is normalized
independently. The solid, $b=0.6$ mm, and dashed, $b=0$, curves represent theoretical
expressions for the penetration depth (right hand scale).}
\end{center}
\end{figure}

The data show, particularly in Fig. 6, that ignoring the profile width, $b=0$, gives a
poor fit. Setting $b=0$ increases the separation of the curves, since it weights the
logarithmic term in $\alpha$ [Eq. (12)] relative to the digamma function, but does not
give the correct curvature as a function of magnetic field.  This confirms that the
density profile is important in determining the EMP mode frequencies.

We also compare Eq. (13) with our data, although $b/h \simeq 0.6$ does not satisfy the
condition $b/h \ll 1$. It is applicable for $L \ll R/h = 10$.  Equation (13), derived
for a semi-infinite plane, yields a constant separation between mode frequencies,
$f\propto L/B$. Aside from a numerical factor, this expression gives a good fit to the
data in the range where it is applicable, namely, $b \gg\ell$ and small values of $L$.
It overestimates the frequency at low magnetic fields due to neglect of the finite
sample radius.  The theoretical curves are adjusted by a numerical factor of order $2$.

The role of the parameter $b$ is shown explicitly in Fig. 7.  The dependence of
frequency on the ratio $|V_{g}|/V_{h}$ at $B=0$ is due to the change in effective sample
radius with a change in guard voltage. The frequency ($\propto 1/R$) changes by $9.5\%$
over the range shown, in agreement with a change in radius of $10\%$.  For the data at
larger fields, $\ell\ll b$, the percent change in frequency with field over the range
shown is $23\%$.  We attribute the difference of $13\%$ to a change in $\lambda$, which
is predicted to be equal to $b$ in this range.  Theoretical frequencies increase by
$15\%$ over the range. We note that the definition of $b$ is somewhat arbitrary, but the
theory curve shown in Fig. 7 is nearly insensitive to the definition of $b$.  We can
match the $23\%$ change in the theory only by a increasing the parameter $b$ by a factor
of $10$. Not only is this adjustment unphysical, but the change is opposite that needed
to improve the fit in Figs. 5 and 6. The decrease in frequency for the lowest value of
$|V_{g}|/V_{h}$ from an extrapolation of the data at higher ratios is due to the
nonlinear dependence of $R$ and $b$ on $|V_{g}|/V_{h}$ for $|V_{g}|/V_{h} < 0.4$.

The lack of any measurable frequency shift upon crossing the melting
curve verifies that the shear modulus has a negligible effect on the
dispersion of the EMP modes.

The magnetic-field dependence of the linewidths is in qualitative agreement with
theoretical values but differs by a large numerical factor.  The data fall below the
theory at large fields. The value of $\tau$ depends on the electric field, proportional
to $n$, pressing the electrons to the surface. The theoretical value of $\tau$ is $7$
$ns$ for $n_{12} = 1.75$ at $400$ mK.  For a comparison with experimental values of
$\tau$, the theoretical value at $400$ mK is $17\%$ less than the experimental value at
a density of $n_{12} = 1.05.$\cite{Mehrotra} This does not explain the quantitative
discrepancy with theory. The predicted linewidth with the theoretical value of the
frequencies is obtained by substituting Eq. (12) into Eq. (14) and is given by $\Delta f
= 1/2\alpha\tau$.  This yields a linewidth that is too large by a factor of
approximately $5$, depending on the field at which experiment and theory are compared,
and it also lies above the data at large fields. The temperature dependence of the
linewidths agrees with that of ripplon-scattering theory.

The experimental variation of the penetration depth with magnetic field is in agreement
with theory for an applicable range of fields. This includes $B > 0.003$ T, which is the
range of validity of Eq. (10), $\omega_c^2 \simeq 5 \omega^2$ for $B = 0.003$ T, and
within which the condition $\lambda\ll R$, assumed in Eq. (15), is satisfied. The
assumption of an exponential decay of the wave amplitude may be violated for $\ell < b$
or $B > 0.0094$ T.  Note that $S(\Delta\omega)^{2}$ is still proportional to the number
of electrons participating in the mode, which is shown in Fig. 10 to vanish at large
fields.

\section{Conclusions}

We find that the dependencies of the EMP mode frequencies on
magnetic field, azimuthal mode number, and the screening parameter
h/R have the general qualitative behavior predicted by Fetter.
Quantitative differences for the lowest modes are small in the range
of application of this theory, which is limited.

The data exhibit the qualitative features of the theory of Volkov and Mikhailov.  The
coefficient $\alpha$ in Eq. (12) overly confines the separation of azimuthal modes and
at low fields leads to a reduction in frequency in contradiction with experimental data.
An adaptation of their theory for a semi-infinite plane to a circular sample fits our
data for large fields and small azimuthal mode number. Our data on mode frequencies
confirm that the penetration depth is determined by the density profile at large fields
as predicted by these authors and Nazin and Shikin. We demonstrate a variation of
frequency with a change of edge density profile.

The magnetic-field dependence of EMP linewidths is in qualitative agreement with theory
to within the scatter in the data.  Experimental linewidths deviate from theory at large
fields. Theory and experiment differ by a numerical factor of $\sim 9$. At large fields
the linewidth is independent of mode number as predicted. The temperature variation of
the linewidth is in agreement with theory.

We are able to extract the dependence of the penetration depth on magnetic field,
assuming an exponential decay of the mode amplitude into the sample.  This assumption
may not apply when the penetration depth is less than the density profile width.  We
show that the number of electrons participating in the EMP wave vanishes at large
fields.  A more exact definition of the profile width would be helpful as well as a
prediction of the decay of the EMP amplitude from the sample edge within the profile
width.

At our largest azimuthal modes, $L = 12$, the wavelength is $\simeq R/2$.  It would be
interesting to investigate higher modes in a sample of larger radius with $\lambda\ll R$
to test theories of the propagation of EMP modes in a semi-infinite
sample.\cite{Fetter2,Mikhailov}

\section{ACKNOWLEDGMENTS}
The authors wish to acknowledge M. I. Dykman, A. L. Fetter and Harsh Mathur for helpful
conversations. This work was supported in part by NSF grant No. DMR-0071622.  M.I.G.
wishes to thank the Turkish Ministry of Education for support.

\end{document}